\documentclass[aps,prl,floats,twocolumn,amsfonts,amssymb,unsortedaddress,floatfix]{revtex4}
\input epsf
\usepackage{bm}
\usepackage{amsmath}
\usepackage{graphicx}
\usepackage{xcolor}
\usepackage{ulem}

\begin{document}

\addtolength{\topmargin}{10pt}

\def\Bbb{\mathbb}

\title{Long-time behavior of  the $\omega \to \alpha$ transition in  shocked Zirconium: Interplay of  nucleation  and plastic deformation}

\author{$^1$Cristiano~Nisoli,  $^1$Honxiang Zong, $^{2,3}$Stephen R. Niezgoda, $^2$Donald W. Brown, $^1$Turab Lookman}
\affiliation{\mbox{$^1$Theoretical Division and $^2$MST Division, Los Alamos National Laboratory, Los Alamos NM 87545 USA} \\
\mbox{$^3$Department of Materials Science and Engineering, The Ohio State University, Columbus OH 43210 USA}}

\date{\today}
\begin{abstract}
We study the thermally activated, slow conversion of the hysteretically retained $\omega$ phase into stable $\alpha$ phase in recovered samples of shocked zirconium. The  $\omega$-phase  decays in time following an algebraic law, unlike the predictions of the nucleation-growth framework for first order transitions, and residual volume fractions of phases and dislocation densities are related by a power law. We propose  an explanation for the annealing mechanism through coupled dynamics of dislocations and phase change. We find that the long-time behavior is controlled by the interplay of dislocations, shear fluctuations, and remnant volume fractions of phases, which lead to an algebraic decay in time. For late time, thermally activated quantities such as the dislocation mobility and nucleation rate set the timescale and control the algebraic behavior, respectively. At high enough temperatures this behavior is effectively indistinguishable from standard Avrami kinetics.  
  
\end{abstract}


\maketitle 

There has been much recent interest in understanding the coupling of phase transformation and deformation processes~\cite{Zong0,Valiev, Bridgman}. The group IV elements and especially Zr and Ti, with relatively easily accessible transition temperatures and pressures, have been excellent  test beds for investigating aspects of deformation and phase transformations under high pressure and shock \cite{Hickman,Jamieson, Vohra, Rabinkin, Kutsar, Greeff, Jyoti, Zong1}. Starting from ambient conditions, these metals undergo an hcp ($\alpha$) to hexagonal ($\omega$) structural transformation under pressure, which on release retains substantial volume fractions of the high pressure $\omega$ phase~\cite{Hickman, Jamieson, Cerreta, Brown2}. The volume fraction of  $\omega$  increases with the peak pressure.  The metastable microstructure of coexisting $\alpha$ and $\omega$ phases in recovered samples,  and the significant hysteresis across the equilibrium phase boundary,   are a reflection of the non group-subgroup nature of the first-order shear and shuffle (phonon) driven transformation \cite{bhattacharya}. In addition, we expect aspects of the slow kinetics and history dependence, affected by defects and heterogeneities, to be consequences of the reconstructive nature of this transformation.

We have previously studied the initial evolution of the retained $\omega$ phase in recovered samples of shocked Zr at several temperatures under isothermal annealing conditions using x-ray diffraction measurements \cite{Zong, Brown}. Our  principal conclusions were that in the temperature range 430-535K, the activation barriers calculated from a modified Kohlrausch-Williams-Watts relation  for  the evolution of the volume fraction for the reverse $\omega \rightarrow \alpha$ transformation increased with the peak shock pressure \cite{Zong}. Molecular dynamics simulations interpreted the changes in activation barrier to be controlled by  hetrogeneous nucleation from  defects, such as dislocations,  in the microstructure. 

The focus of the current work is to develop a phenomenological model for the {\it long time behavior} that describes the annealing via a coupled dynamics of dislocations and phase change. We derive an algebraic dependence for the late time evolution of the volume faction of phases at relatively low temperatures, which in the high temperature regime recovers the standard Avrami kinetics.
While the kinetics of high temperature recrystallization is generally well understood~\cite{Humphreys} and found to follow the established Avrami type sigmoidal profile typical of nucleation and growth processes~\cite{Avrami}, we show here that slow aging of the retained phase in zirconium at low temperatures involves a more subtle interplay with thermally activated dislocations. This is a mechanism reminiscent of shear-driven martensitic transformations. In the case of transformation induced plasticity (TRIP) it leads to a unique combination of high strength and ductility ~\cite{Patel, Fisher} in steels. Localization of externally applied deformation, shear banding,  creates potent sites inside the disappearing phase, on which the new phase can nucleate~\cite{Patel, Fisher, Olson, Stringfellow, Perlade}. Here we show that even in the absence of external shear, the thermally activated motion of dislocations can determine fluctuations of plastic shear capable of inducing a martensitic transition that eliminates a metastable phase. 

We introduce a  simplified model of how this can come about, and we test this framework on experimental data obtained by thermal annealing of recovered shocked  Zirconium (hcp $\alpha$-phase:  $P63/mmc$,  $c/a = 1.593$ ; simple hexagonal $\omega$-phase: $P6/mmm$, $c/a = 0.623$).
\begin{figure*}[t!!!]
\begin{center}
\includegraphics[width=.9\columnwidth]{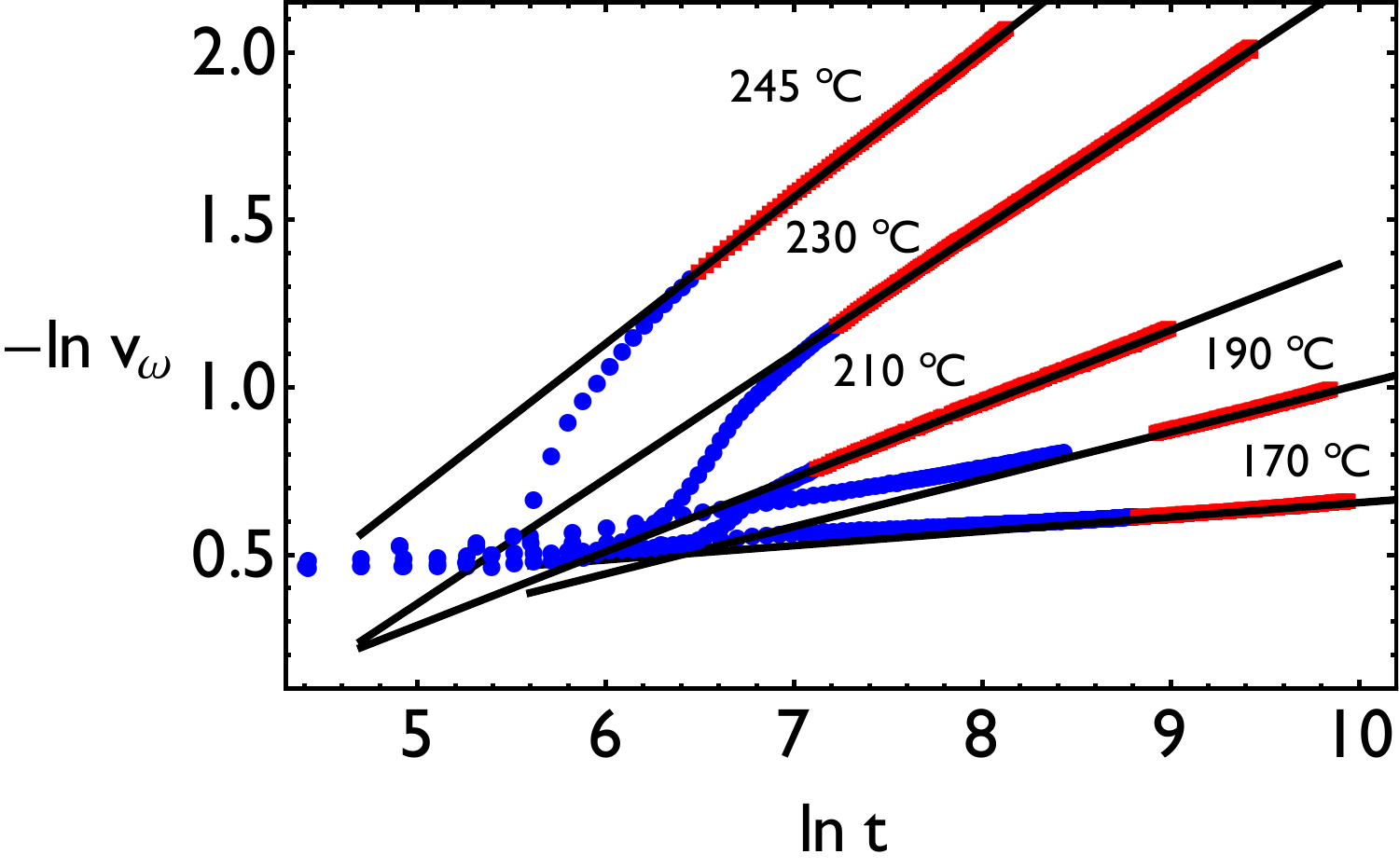}\hspace{10 mm} \includegraphics[width=.848\columnwidth]{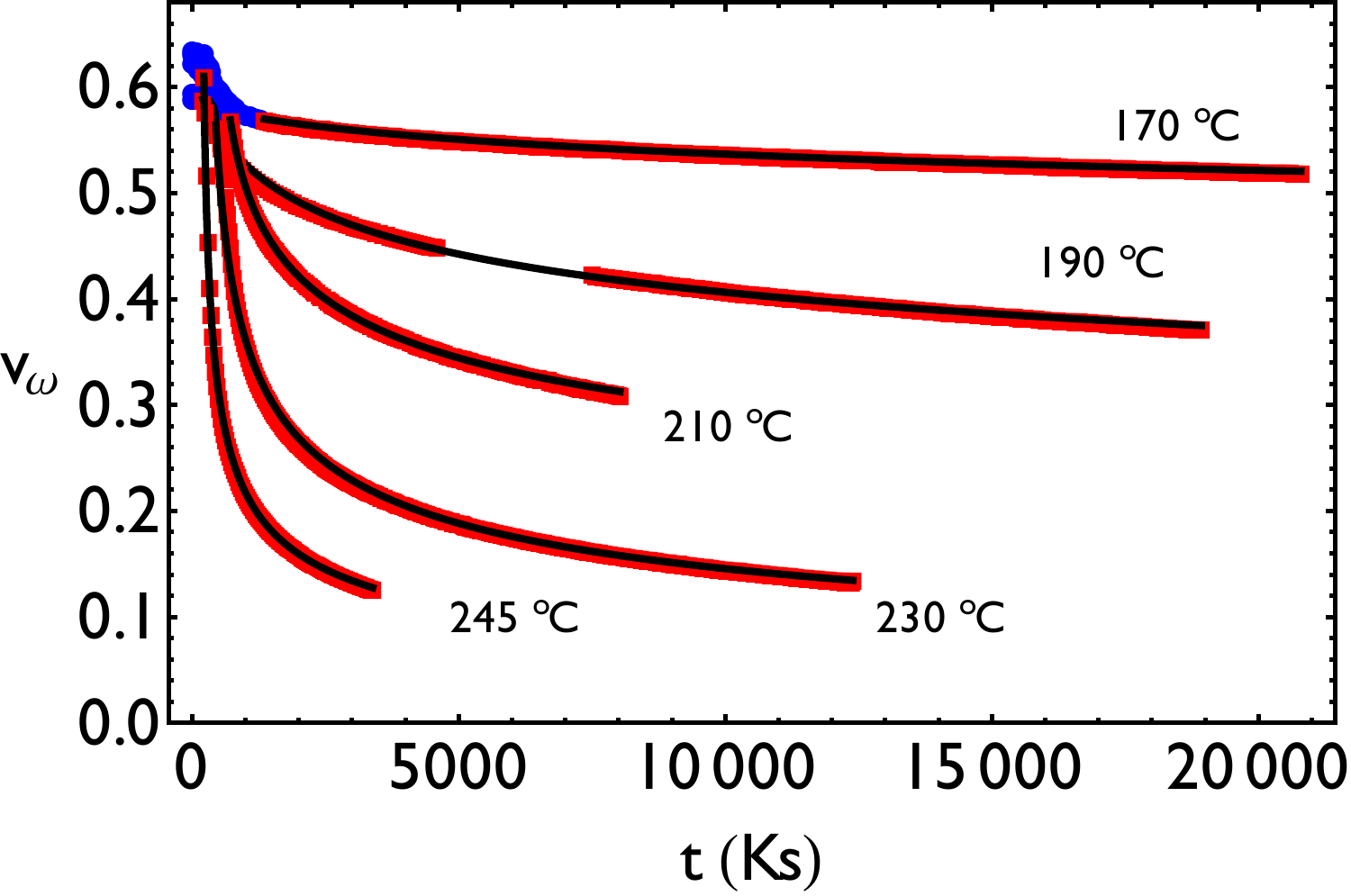}
\includegraphics[width=.9\columnwidth]{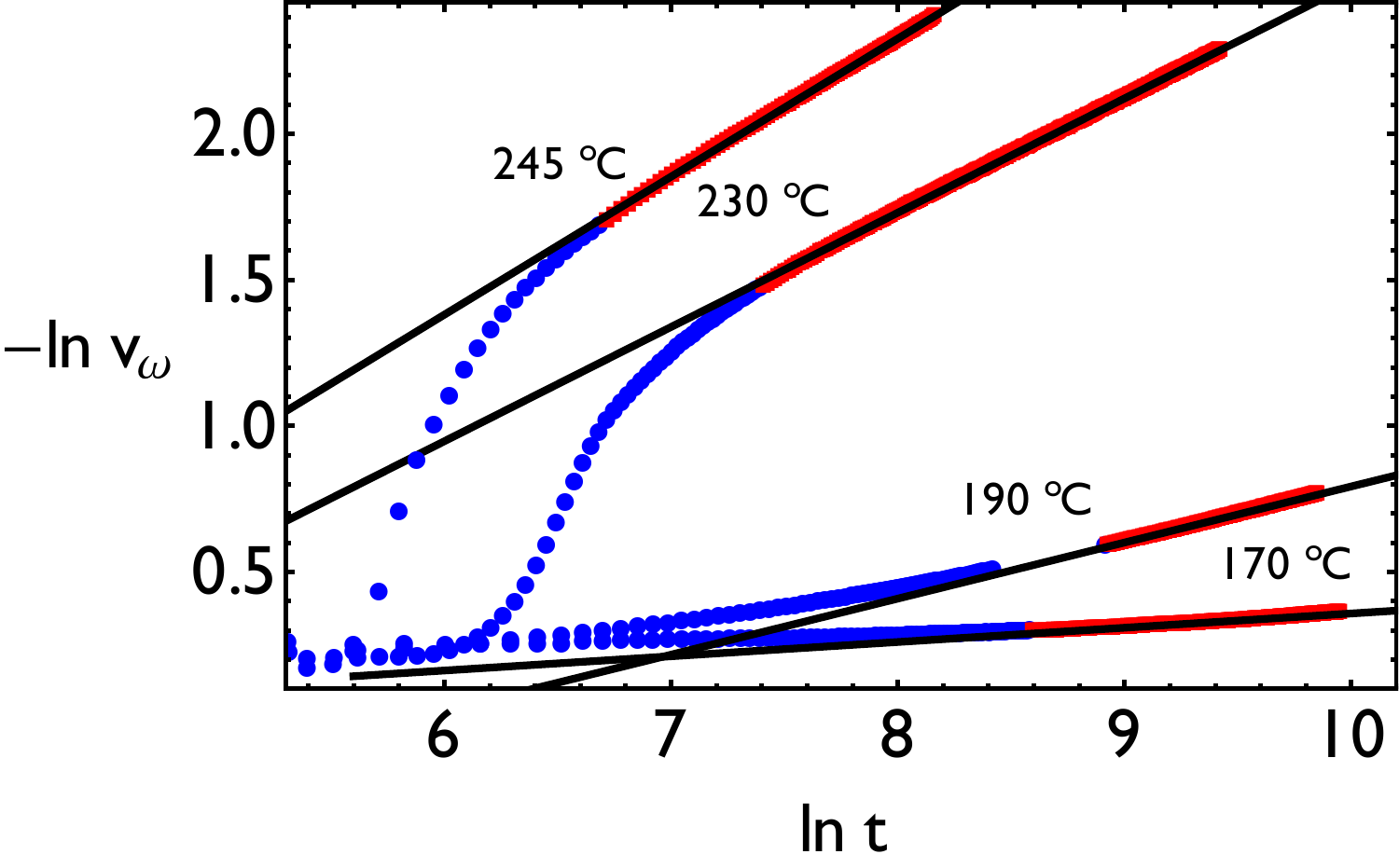}\hspace{10 mm} \includegraphics[width=.848\columnwidth]{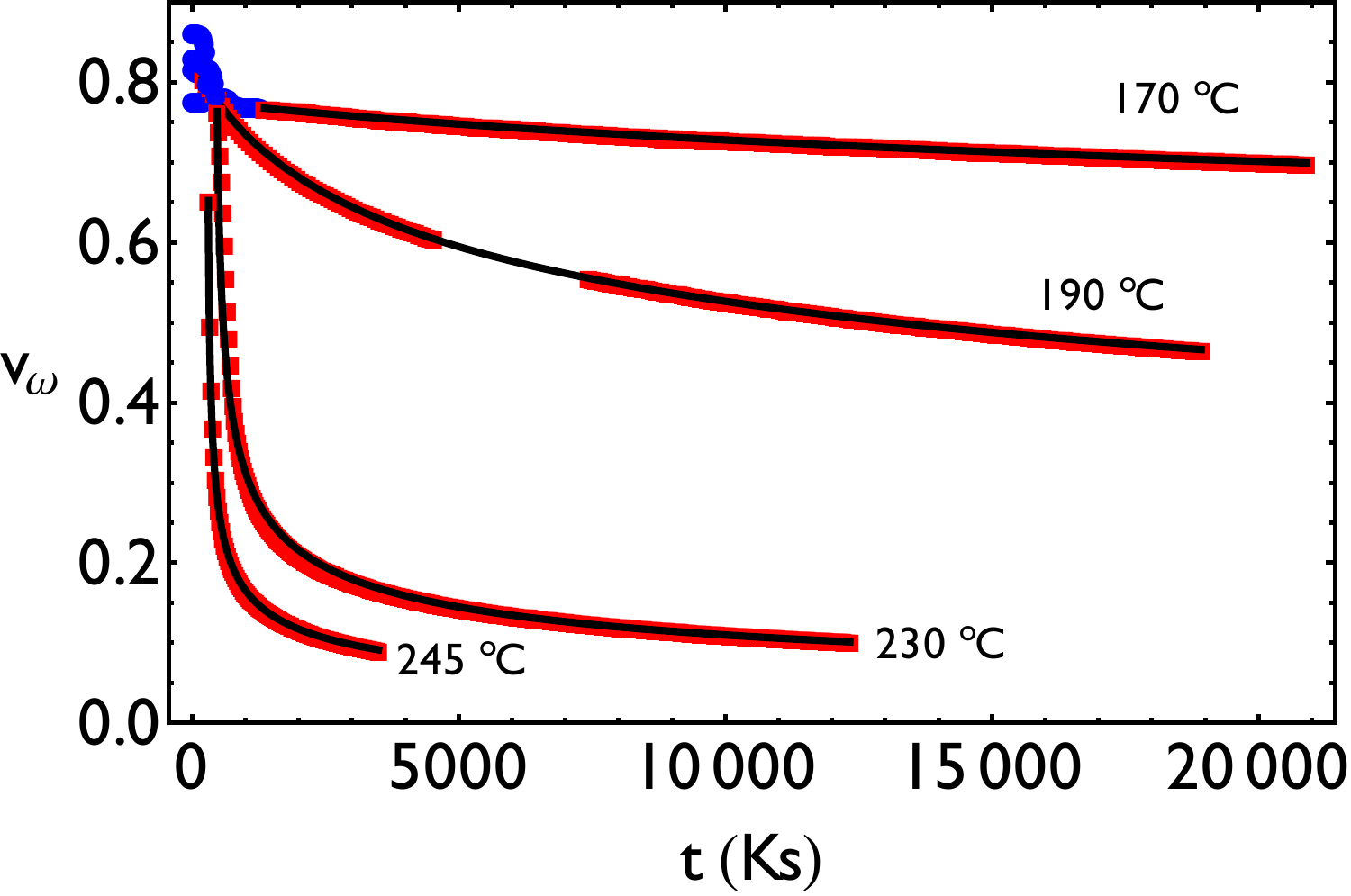}
\caption{Fits  to  volume fraction of the disappearing $\omega$-phase obtained from experimental data taken from the annealing of  samples already shocked at 8GPa (top left) and  at 10 GPa (bottom left) using Eq.~(\ref{vt}). The left panels of  log-log plots show  power law behavior  at long times. Only data pertaining to long time behavior (red dots) are fitted.}
\label{v_t}
\end{center}
\end{figure*}
Polycrystalline $\alpha$-Zr  shocked loaded to 8 GPa and 10.5 GPa  results in a retained volume fraction of $\sim60\%$ and $\sim80\%$ respectively of the metastable {$\omega$-phase} ~\cite{Brown}. When the recovered shocked samples  are heated, the  data show a slow annealing as well as  monotonic relationships between the density of dislocations and the volume fraction of the remnant phase~\cite{Brown}. Transitions of this kind are often described by the Johnson-Mehl-Avrami-Kolmogorov approach~\cite{Avrami}. However the standard Avrami test on the data shows, at large times, strong deviations from  uniform nucleation and growth behavior and its typical sigmoidal behavior. 

This data, previously reported in the literature \cite{Zong, Brown}, is plotted  in Fig. 1. We first recognize that  at long times the decay is algebraic in time and is well fitted by 
\begin{equation}
 v_{\omega}=\frac{v_{\omega,0}}{(1+  t/\tau)^{\chi}}.
\label{vt}
\end{equation}
where $v_{\omega}$ is the volume fraction of the disappearing {$\omega$-phase}, and the exponent $\chi$ appears to be thermally activated. The key to understanding this algebraic behavior, which is one of our results which we will derive in this work,  lies in combining plasticity with nucleation of the new phase.  We assume during shock that the material first twins and then nucleates dislocations, which can reach a considerable density before and possibly while the material finally undergoes a  transformation toward the $\omega$-phase. These dislocations, may be interpreted as  a stabilizing network allowing for  retention of the  metastable $\omega$-phase. Then, as the sample is heated,  the acquired mobility of these thermally activated dislocations can undermine stability, thus providing the driving force for the transformation.

Indeed, martensitic transformations are mediated by a shear strain~\cite{Patel, Fisher}, for example,  the retained, metastable austenite phase in TRIP-steel, and its martensitic shear driven transformation~\cite{Olson, Stringfellow, Perlade}. 

Following Stringfellow {\it et al.}~\cite{ Stringfellow}, we  write the proportionality 
\begin{equation}
\dot  v_{\omega}\propto -  v_{\omega} \dot \gamma,
\label{string}
\end{equation}
relating volume fraction to strain rate $\dot \gamma$ in the  { shear-induced} nucleation on potent sites created by plastic strain~\cite{Perlade}. Stringfellow et al. observed that for TRIP steels the strain-induced nucleation occurs at shear band intersections, and included a proportionality constant to accounted for the nucleation rate of shear bands at low strains. Here the proportionality constant will be  related to the temperature dependent dislocation removal rate. In the case, e.g., of TRIP steel, the shear is  global and externally applied. In our system, however, the total strain rate is endogenous, and ambient. In particular $\dot \gamma$ is not a total derivative,  but rather the average of the absolute value of the local strain rates, and thus, in general $\gamma\ne\int\dot \gamma dt$.

The experimental data provides average and coarse information on dislocation densities, which can be obtained from averaging the peak widths of the X-ray data~\cite{Brown} .We can relate such information to the local shear rates. Indeed, Kocks and Mecking~\cite{Kocks, Teodosiu} have introduced a phenomenological equation to  relate variations in the  shear rate to the density of dislocations:
\begin{equation}
d\rho/d\gamma=c_1 \sqrt{\rho} -c_2 \rho.
\label{KM}
\end{equation}
Equation  ({\ref{KM}) describes the storage (first term) and annihilation (second term) of dislocations in a material subjected to environmental shear, in our case  coming from the collective motion of dislocations (plastic shear) and from phase transformation (transformation shear). 

 \begin{figure}[t!]
\begin{center}
\includegraphics[width=.9\columnwidth]{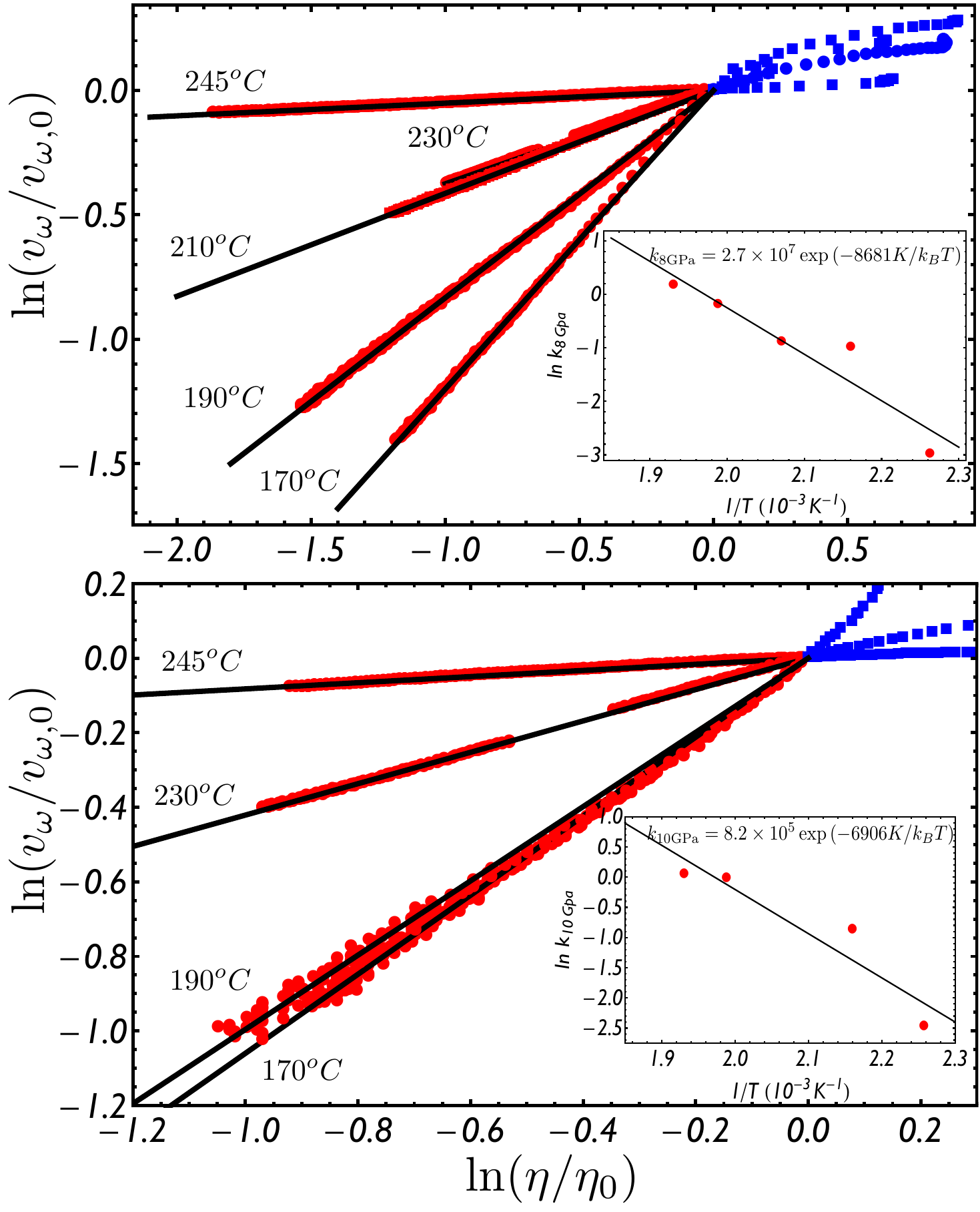}
\caption{ Eq.~(\ref{vrho2}) reveals the power law behavior for the volume fraction and dislocation density parameter $\eta=\rho-\rho_{\infty}$ for zirconium shocked at 8 GPa (Top, at temperatures of 170, 190, 210, 230, and 245 C) and 10 Gpa (bottom, at temperatures of 170, 190, 230, and 245 C); data in red are fitted, data in blue are not fitted, as in Fig.~1. Insets: The fitting exponent $k$, which is the nucleation rate for Eq.~(\ref{munu2}) obtained from  fits to Fig.~\ref{dis}, is a thermally activated nucleation rate; an Arrhenius fit returns  higher  activation energies for $\alpha$-phase nucleation in  the samples shocked at the lower pressure.}
\label{dis} 
\end{center}
\end{figure}

In general, the constant $c_2$ is proportional to the critical annihilation distance for dislocations; the term $c_1$, is the subject of much recent research and its dependence on the specificity of the problem  is still debated~\cite{Devincre}. We can neglect these issues by noting that $c_1$  simply defines the stationary density of dislocations $\rho_{\infty}=(c_1/c_2)^2$ at which, in (\ref{KM}), $d\rho/d\gamma=0$ because of our focus on the asymptotic behavior. We can then eliminate $c_1$ by introducing  $\eta=\rho-\rho_{\infty}$ and expanding the Kocks-Mecking equation around $ \eta \sim 0^+$, thus obtaining in the asymptotic regime $d \eta/d\gamma= -c_2 \eta/2$,
 or equivalently
\begin{equation}
\dot \gamma=-\frac{2}{c_2}\frac{\dot \eta}{ \eta}~.
 \label{gamma}
\end{equation}
Thus from  Eq.~(\ref{string}) and  Eq.~(\ref{gamma}) it is immediate to obtain
\begin{equation}
\frac{d v_{\omega}}{ v_{\omega}}=k \frac{d \eta}{\eta},
\label{munu2}
\end{equation}
which establishes a   proportionality between the relative rates of variation of the dislocation density and volume fraction during phase change (here $k$ is a nucleation rate). 

Fortunately this  interesting  result can be tested using experimental data. Indeed, from  Eq.~(\ref{munu2}) we can now obtain  a power law relating  volume fraction and   dislocation density
\begin{equation}
\frac{v_{\omega}}{v_{\omega,0}}=\left( \frac{\eta}{\eta_0}\right)^k,
\label{vrho2}
\end{equation}
where the exponent is the nucleation rate $k$, $\eta_0=\rho_0-\rho_{\infty}$ is the density of mobile dislocations at the beginning of the process, and $v_{\omega,0}$ is the initial volume fraction. 

In  Fig.~\ref{dis} we we demonstrate the power law behavior of  Eq.~(\ref{vrho2}) via a logarithmic plot which is used to fit the constant $k$: $\rho_0$ is experimentally given, while $\rho_{\infty}$ has to be fitted, as the annealing is never complete in these experiments. In the left panels we demonstrate the power law of  Eq.~(\ref{vrho2}) via a log-log plot which is used to fit the constant $k$ at large times. Since $k$ is a nucleation constant, we expect it to be thermally activated, or
\begin{equation}
k\propto\exp(-\Delta W/k_BT),
\label{activated}
\end{equation}
with activation energy  $\Delta W$. 

Indeed Fig.~\ref{dis} shows the Arrhenius fit for $k$ which returns the values for the activation temperature $T_{\mathrm{8GPa}}=8.7\times 10^3K\simeq0.75~eV$ and $T_{\mathrm{10GPa}}=6.9 \times10^3K\simeq 0.6~eV$, consistent with the results of  atomistic calculations~\cite{Zong}. The fact that activation energies for nucleation rates are different in the two cases points to an inherent difference in the microscopic structure of samples shocked at different pressures,  in accordance with our previous findings~\cite{Zong}. 
%
%
%
%

%
\begin{figure*}[t!]
\begin{center}
\includegraphics[width=.847\columnwidth]{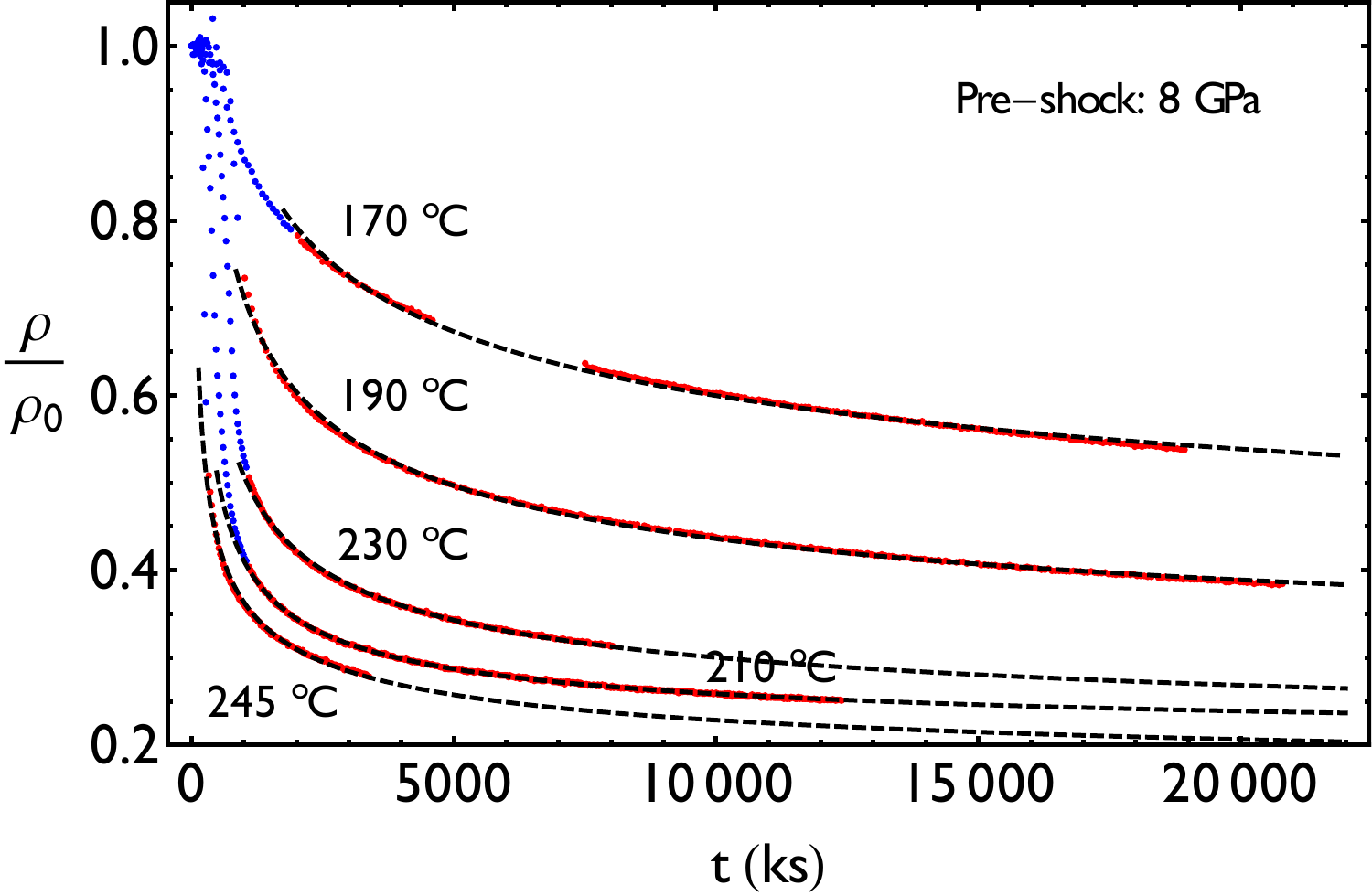}\hspace{10mm}\includegraphics[width=.85\columnwidth]{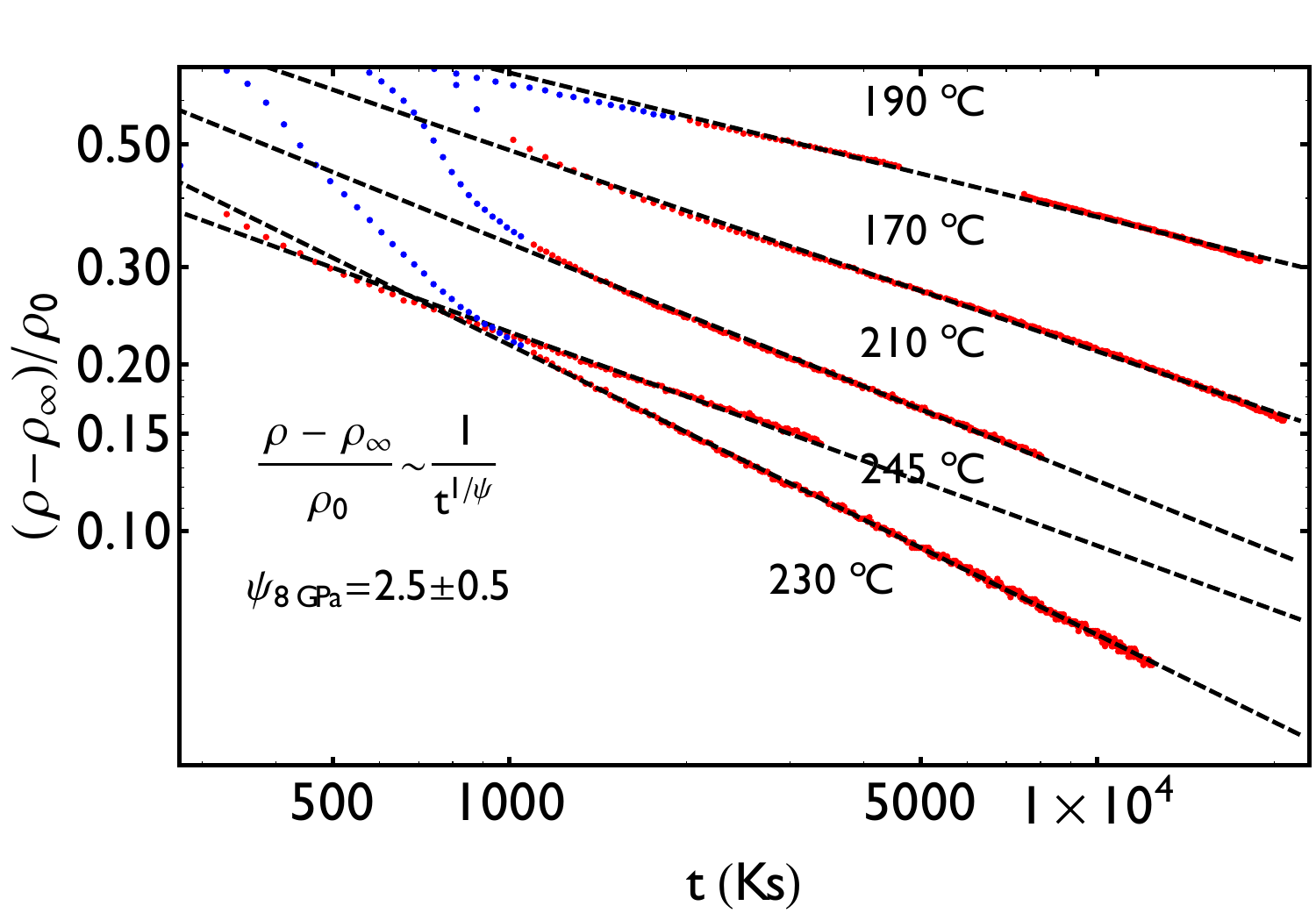}
\includegraphics[width=.847\columnwidth]{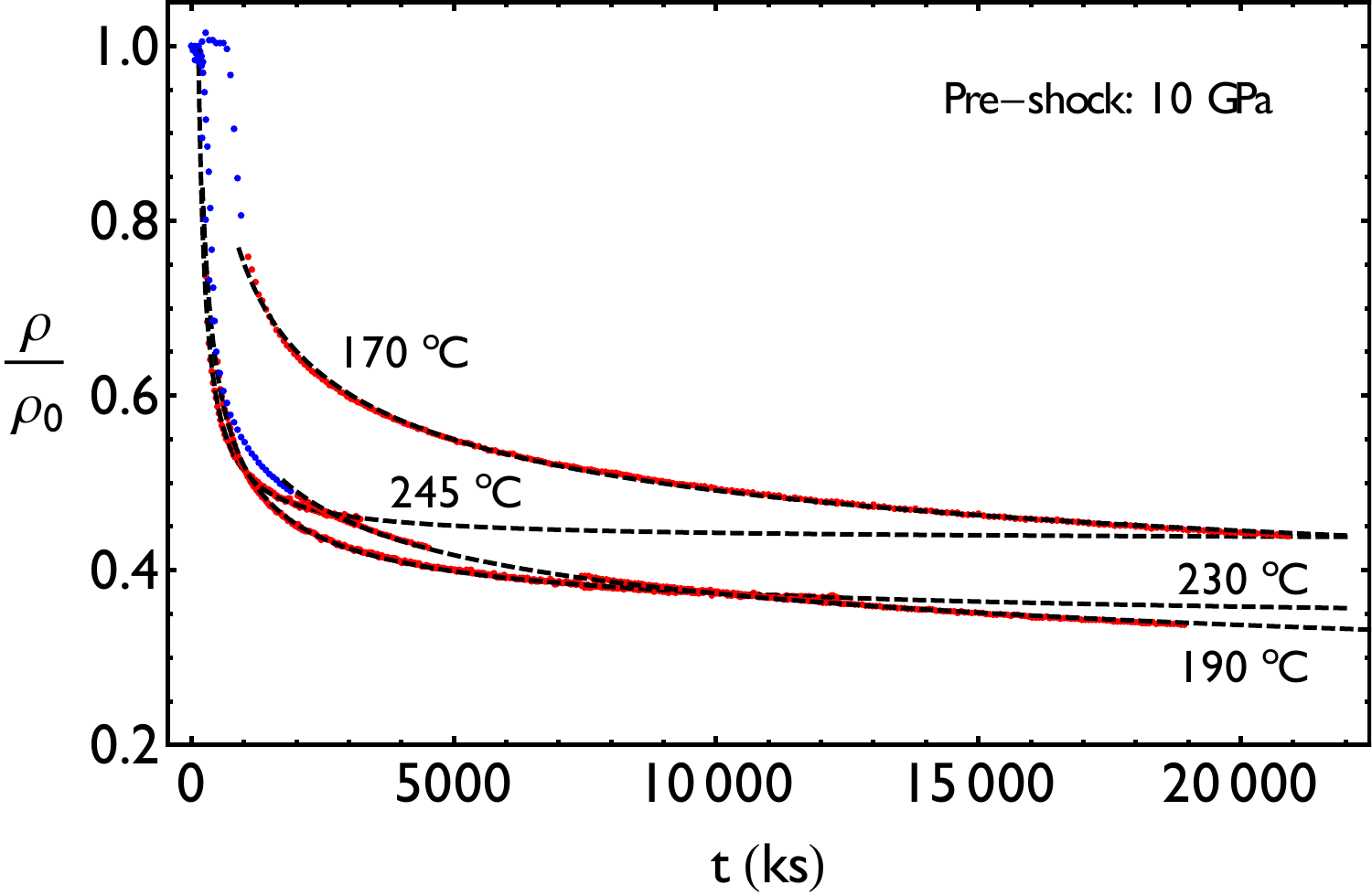}\hspace{10mm}\includegraphics[width=.85\columnwidth]{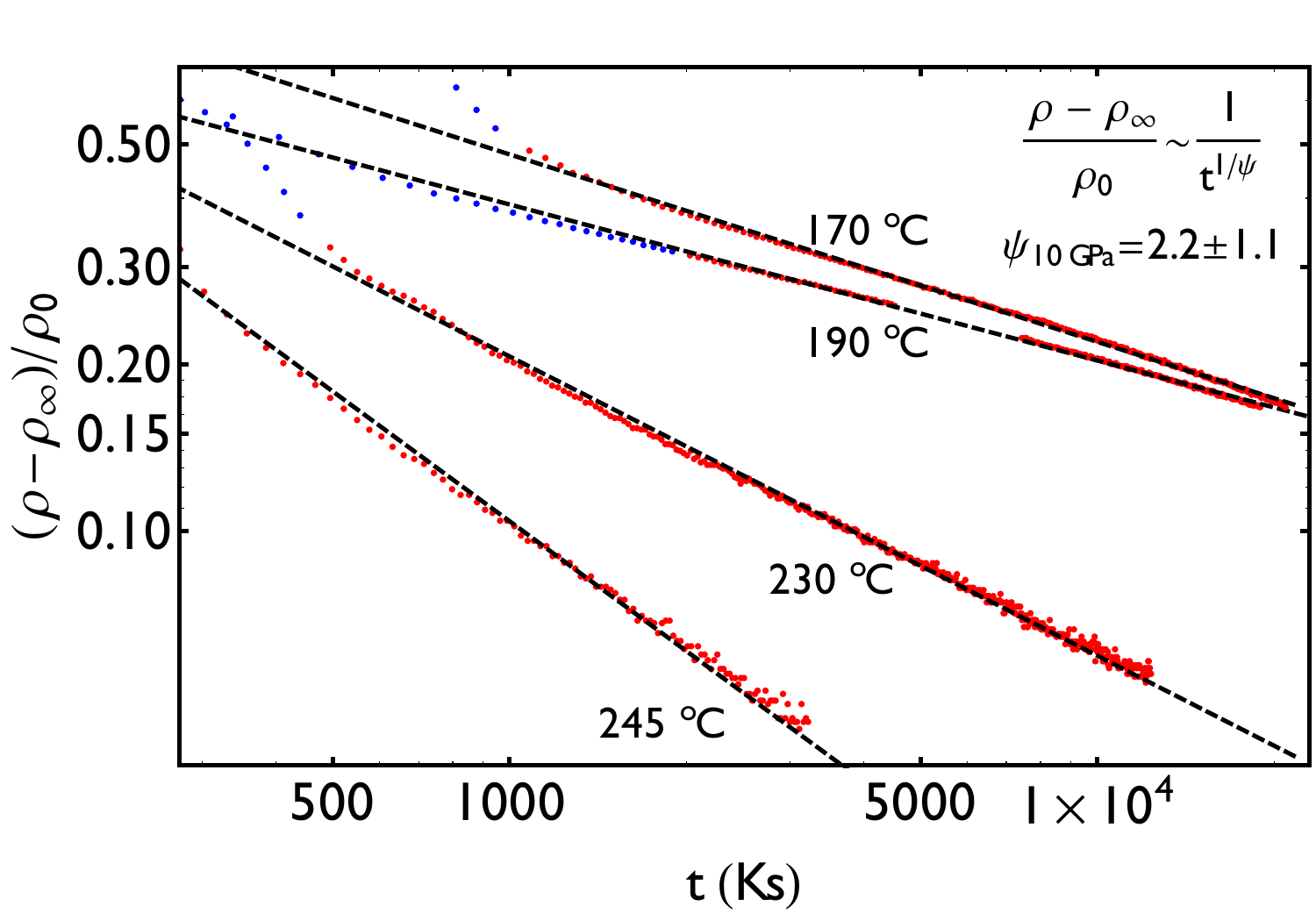}
\caption{Fits to Eq.~(10) for experimental data of dislocation densities normalized at the initial value  as a function of time for samples annealed at different temperatures  and schocked at 8 GPa (top panels) and10 GPa (bottom panels). Log-log plots (right panels) show the algebraic decay from Eq.~(\ref{r}).  The fitted remnant density of dislocations $\rho_{\infty}$ appears scattered amongst samples and temperatures, and is possibly sample dependent~\cite{Brown}; it is however not a relevant parameter in our theory.  Data in red are fitted, data in blue are not fitted.}
\label{dis_t}
\end{center}
\end{figure*}

To deduce the time evolution   of the normalized volume fraction $ v_{\omega}(t)$ it is sufficient to know the time dependence of the dislocation density $\rho(t)$. This can be resolved in a coupled dynamics that involves plasticity and phase change. There are two sources of shear: transformation shear $\gamma_t$ and plastic shear $\gamma_p$.
The source of plastic shear rate $\dot \gamma_p$ is the motion of the dislocations in the $\omega$-phase, and is well known from the Orowan equation~\cite{Orowan}
\begin{equation}
\dot \gamma_p =b \rho_m w_d,
\label{Orowan}
\end{equation}
in which $w_d$ is the thermally activated mobility of the mobile dislocations $\rho_m$, $b$ is the magnitude of the Burgers vector. 
In our phenomenological approach   $\dot \gamma$ is not a total derivative, as explained above when introducing the  Kocks-Mecking equation, but rather the  coarse grained shear,  over a sufficiently large region of  the {\it absolute value} of the intensity of the local shear rates.   Furthermore the relationship between mobile dislocations $\rho_m$ and the excess dislocations $\eta$ is unknown to us. In general the mobile dislocations will be less than $\eta$ as some of the annihilating dislocations might be non-mobile. In the absence of more precise information, {we proceed with the reasonable ansatz} that the {\it relative} decrements of mobile dislocations and of dislocations tout-court are proportional, or $d\rho_m/\rho_m=\psi d\eta/\eta$. This, together with the Orowan equation (\ref{Orowan}) leads us to write
\begin{equation}
\dot \gamma_p =c_3 \eta^{\psi},
\label{Orowan2}
\end{equation}
with $\psi>1$ and $c_3$ proportional to the dislocation mobility $w_d$.

The other source of shear, the transformation shear $\gamma_t$, is associated to the structural change in unit cells of the $\alpha$ and $\omega$ phases. Clearly, its rate is  proportional to the rate of phase change, or
%
$\dot \gamma_t =c_4 \dot v_{\omega}$,
where $v_{\omega}$ is the volume fraction of the $\omega$ phase, and $c_4$ depends on the geometric parameters of the unit cells in the two phases. Indeed from  Eq.~(\ref{string}), we have that in the asymptotic limit  most of the shear rate is purely plastic, or
%
 ${\dot \gamma_t}/{\dot \gamma}\propto  v_{\omega}\to 0$.
%
This allows us to  approximate $\gamma_p \simeq \gamma$ in  Eq.~(\ref{Orowan2}). Then,
in the approximation of long times, $\rho(t)$ can be found from ~(\ref{Orowan2}) and~(\ref{gamma}) to be
\begin{equation}
\eta=\frac{\eta_0}{(1+t/\tau)^{1/\psi}}.
\label{r}
\end{equation}
In Fig.~\ref{dis_t},  we demonstrate  that Eq. (\ref{r}) fits the data very well.  Equation ~(\ref{r}) describes a slow, algebraic decline at long times, controlled by the parameter  $\tau^{-1} \propto c_3\propto w_d$,  which depends on temperature through the thermally activated mobility of dislocations $w_d$.  Interestingly,  the decay is faster at higher temperatures (because $\tau$ of Eq.~(\ref{r}) depends on the activated dislocation mobility) and for samples pre-shocked at higher pressure (because more dislocations have been nucleated to begin with). Yet   the  value of the exponent $\psi$ from Eq. (\ref{Orowan2}), which represents the ratio of relative decrement between mobile dislocations and dislocations proper is {\it almost} the same in all cases and close to the value of $\psi=2.5$.

Equation (\ref{vt}) can now finally be deduced from (\ref{vrho2}) and (\ref{r}) which imply
%
$\chi=k/\psi$:
the exponent $\chi$, which controls the lond time algebraic decay of the remnant volume fraction,  is thus also, like $k$, an activated quantity that follows an Arrhenius law. 
This result suggests the following: {\it  while the dislocation mobility $w_d$ sets the timescale, the nucleation rate $k$ controls the algebraic decay of the residual volume fraction of the $\omega$-phase}. As both quantities are thermally activated, higher temperature implies not only a smaller timescale (controlled by $w_d$) but also a larger relative decrement for the same relative increment of time (controlled by $k$).


Our work elucidates the limits of the standard nucleation and growth framework~\cite{Avrami} which can be relevant to other systems. We have seen that the algebraic approach to equilibrium is a consequence of the coupled dynamics of  phase transformation and dislocations. 
When the temperature is high enough, the conventional Avrami formalism can still provide a reasonable description. Let us see how that can come about. 

From  Eq.~(\ref{vt}) we can write directly
\begin{equation}
\frac{\dot v_{\omega}}{v_{\omega}}=-\frac{k}{\psi \tau} \left(\frac{v_{\omega}}{v_{\omega,0}} \right)^{\frac{\psi}{k}},
\label{vt2}
\end{equation}
which shows that the total nucleation rate is not constant, as in usual nucleation and growth, but rather a power law of the concentration of the disappearing phase. However, at high temperatures, $k$ becomes exponentially large and $\left({v_{\omega}}/{v_{\omega,0}} \right)^{{\psi}/{k}}\simeq1$ unless  unless ${v_{\omega}}/{v_{\omega,0}}$ is very small, say below the experimental error. Then  the nucleation rate can be well approximated by the constant value $k/\psi \tau$. Of course, mathematically, the behavior at long times will always become algebraic. However, for all practical purposes at large enough temperatures, that might happen only  when the residual volume fraction is as  low as to be negligible or unmeasurable, while in general  Eq.~(\ref{vt2}) shows that the measured nucleation rate might in fact appear constant at any experimentally reasonable value of the volume fraction. 

In the context of the behavior of $v_{\omega}$ vs $\eta$, described by   Eq.~(\ref{vrho2}), at high temperature and thus large  $k $ corresponds  a sudden drop in volume fraction on a  timescale $\tau \psi/k$, while the density of dislocations would show little change, its timescale being $\tau$; thus   a standard nucleation  approach can be effectively regained when the two dynamics, for phase transformation and dislocations, become decoupled.

In conclusion, we show that a coupled dynamics of plasticity and nucleation leads to an algebraic transformation law in which dislocation mobility defines the timescale whereas the nucleation rate on potent sites controls the power law behavior. Beside proposing a possible explanation for the slow aging of the metastable $\omega$-phase in Zirconium, we are also elucidating one of the mechanisms by which, more generally,  the accepted nucleation and growth framework can  mask a more complex dynamics, one that becomes apparent only at lower temperatures. This is most likely not an isolated case and similar mechanisms may be at work in the aging of metastable multiphase materials under extreme thermodynamic and boundary conditions, such as at nanoscale or in film geometries.

This work was carried out under the auspices of the National Nuclear Security Administration of the U.S. Department of Energy at Los Alamos National Laboratory under Contract No.~DEAC52-06NA25396.

\end{document}